\documentclass[pre]{revtex4}

\usepackage{graphicx,subfigure}
\usepackage{bm}
\usepackage{multirow}

\begin{document}

\preprint{LAUR-06-5040}

\title{The functional structure of cortical neuronal networks grown {\it in vitro} }

\author{Lu{\'i}s M. A. Bettencourt}
\email{lmbett@lanl.gov}
 \affiliation{T-7, Theoretical Division,  MS B284, Los Alamos National Laboratory, Los Alamos NM 87545 }

\author{Greg J. Stephens}
\email{gstephen@princeton.edu}
\affiliation{Lewis-Sigler Institute for Integrative Genomics and Center for Study of Brain, Mind and Behavior, Princeton University, Princeton  NJ 08544}

\author{Michael I. Ham, and Guenter W. Gross}
\affiliation{Center for Network Neuroscience, University of North Texas, Dept. of Biological Sciences P.O. Box 305220 Denton, TX 76203-5220}

\date{\today}

\begin{abstract}
We apply an information theoretic treatment of action potential time series measured with microelectrode arrays to estimate the connectivity of mammalian neuronal cell assemblies grown {\it in vitro}.   We infer connectivity between two neurons  via the measurement of the mutual information between their spike trains.  In addition we measure higher point multi-informations between any two spike trains conditional on the activity of a third cell, as a means to identify and distinguish classes of functional connectivity among three neurons.   The use of a conditional three-cell measure removes some interpretational shortcomings of the pairwise mutual information and sheds light into the functional connectivity arrangements of any three cells.  We analyze the resultant  connectivity graphs in light of other complex networks and demonstrate that, despite their {\it ex vivo} development, the connectivity maps derived from cultured neural assemblies 
are similar to other biological networks and display nontrivial structure in clustering coefficient, network diameter and assortative mixing.  Specifically we show that these networks are weakly disassortative small world graphs, which differ significantly in their structure from randomized graphs with the same degree.   We expect our analysis to be useful in identifying the computational motifs of a wide variety of complex networks, derived from time series data.
\end{abstract}

\maketitle

\section{Introduction}
\label{sec:intro}

Understanding and quantifying the dynamical mechanisms used by the nervous system to store and process information remains one of the greatest challenges to contemporary science. At present, broad outlines of the physical mechanisms that underpin the basic functioning of single neurons and synapses in the brain are  understood \cite{BookNeuroscience}. However, this detailed knowledge of individual units sheds little light into the origin of  the unmatched computational power of mammalian nervous systems, achieved despite characteristic operating times that are  six orders of magnitude slower than those of modern digital computers.

The computational nature of the brain lies therefore principally in the ensemble properties of neurons, synapses and their emergent complex, dynamical networks. Over the last few years the interaction structure of many complex systems has been mapped in terms of graphs, which can in turn be characterized using tools of  statistical physics \cite{Barabasi_2002}. This approach has revealed broad classes of networks such as {\it small world graphs} \cite{Watts} and {\it scale free networks} \cite{ScaleFreeNets}, which occur across fields of study, from technological networks, such as the internet, to various biological and social systems. The structural properties of these graphs, such as their degree distribution or their local transitivity, moreover, have been suggested to result from optimization constraints \cite{BookNetworks1,BookNetworks2} or network growth dynamics \cite{BookNetworks2}, thus connecting graph structure to operational definitions of function, independent of a system's details. These lines of research provide new quantitative insights, connecting the interaction structure of a complex systems to novel definitions of function in fields where quantitative syntheses have been lacking. 

The mammalian  nervous system is, arguably, simultaneously the most complex and fascinating of all networks. Studies that pursue its quantitative understanding in terms of structural, dynamical  networks of simple units (neurons and synapses) \cite{fMRI} are now beginning to be possible.  
Notwithstanding recent reconstructions for invertebrate cells \cite{Shefi_PRE_2002_MorphNeuralNets}, which form relatively small numbers of connections to other cells, measuring the structure of living networks of neurons directly is a difficult problem due to typical huge synaptic densities (about $10^8$ per mm$^3$ in cortex) and large potential connectivity, with degree per cell of order $10^4$ \cite{ChainsMemory1}.  Much more accessible are electrophysiological signals - action potentials or 'spikes' - generated by neurons. The electro-physiological technology to measure these signals over hundreds or even thousands of individual cells is now mature. Other new techniques based on fluorescence of electrically active cells can also be used to map correlated activity, and putative connectivity \cite{Jia_PRL_2004_FlourescentNets}. Because action potentials are the signals encoding information in the brain, the study of collective network activity in terms of their time series is both natural and important.  From these types of data  'networks of information' can be inferred, revealing how signals are shared between neurons, stored and collectively processed.

Here we construct networks of neurons, based on the information theoretic treatment of time series of action potentials from cortical neuronal networks grown {\it in vitro}. Though these systems may differ substantially from an intact brain they are interesting both as non-trivial starting points for {\it in vivo} analysis, and as examples of living neuronal networks capable of non-trivial computation  \cite{Marom_QRevBio_2002_Review}. For these reasons neuronal networks {\it in vitro} have received much recent attention in terms of their collective dynamical and statistical properties \cite{Segev_2001,Tateno_2002,Segev_2002,Beggs_2003,Segev_2004,Beggs_2004,Hulata_2004,Beggs_2005}. 

Below we develop a methodology, based on the information theoretic treatment of multiple cell spike time series, to map the structure of networks of neurons. We show that the resulting graph displays non-trivial structure shared by other complex networks. Our approach also identifies collective functional modules consisting of  groups of neurons that enable different modes of information processing in nervous  systems.

The remaining of this paper is organized as follows. In section~\ref{sec:experiments} we give details of our experimental setting and of the electrophysiological activity recordings. We also introduce  the several information theoretic quantities to be used subsequently to reconstruct a proxy connectivity network. We show how the use of several correlation and information theoretic quantities taken together can lead to the identification of specific connectivity structures, when multiple neurons are considered simultaneously. In section~\ref{sec:info-theory} we give the results of these methods applied to data. Section~\ref{sec:network} analyzes the resulting graph in light of the properties of other complex networks. Finally we present the outlook for the methods proposed here, particularly in determining the dynamical evolution of networks and their collective response under driven conditions. In addition in Appendix A we prove  a relation between definitions of information theoretic quantities, introduced in section ~\ref{sec:info-theory}. 

\section{Experiments and data}
\label{sec:experiments}
 
We analyze the time series of mammalian (murine) spiking cortical networks living in culture over microelectrode arrays \cite{Gross_1994,Keefer_2001,GrossGopal_2006}. These platforms, along with other types of neuronal culture \cite{Beggs_2003,Beggs_2004,Beggs_2005},  currently allow the most extensive coverage of network elecrophysiological activity, in terms of both numbers of cells and length of recordings.  Naturally, cultured cell networks have presumably a simpler structure than their counterparts {\it in vivo}. Nevertheless they are fully active neuronal networks, displaying many of the qualitative properties of developing mammalian tissue \cite{Marom_QRevBio_2002_Review}. The topology of networks of neurons formed {\it in vitro} remains largely uncharted, although recent work in this direction has started to appear \cite{Shefi_PRE_2002_MorphNeuralNets,Jia_PRL_2004_FlourescentNets}. 

\begin{figure}
 \begin{minipage}[b]{0.9\linewidth}
 \includegraphics[angle=0,width=4.5in]{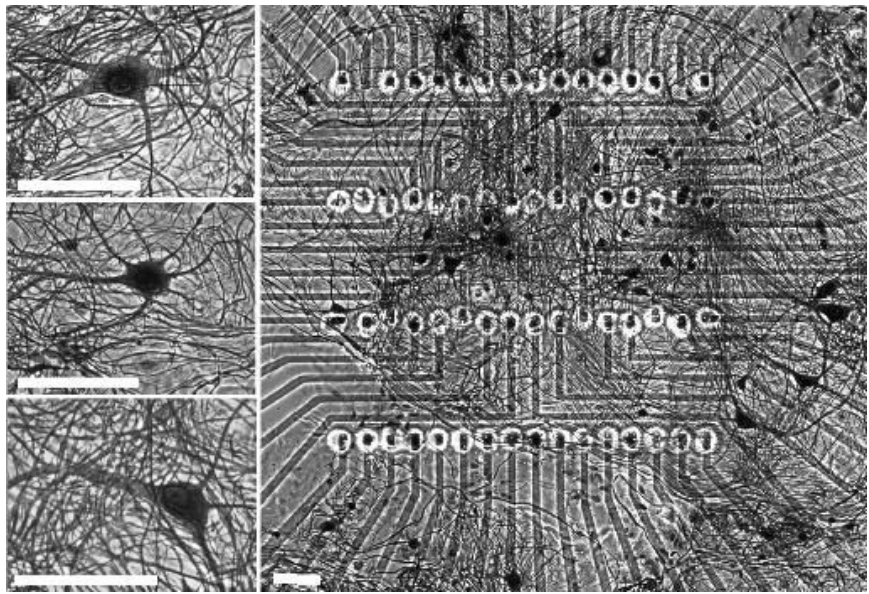}
\end{minipage}
\caption{A neuronal culture (right) over a microelectrode array. The uniform grid of points shows electrodes where neuronal activity is measured. Left panels show individual neurons. White bars are 50 $\mu m$  (center-to-center distances between electrodes is 40 $\mu m$).} 
\label{fig:culture}
\end{figure}

Neuronal networks on microelectrode arrays (MEAs) are assemblies of mammalian nerve cells growing on fields of non-invasive, substrate-integrated  microelectrodes, see Fig.~\ref{fig:culture}.  The cells remain viable for over six months and form spontaneously active networks. Tissue is dissected from mouse embryos, dissociated, and then seeded on MEAs.  Neural activity takes the form of action potentials (or spikes): short pulses  (at which we record a time stamp) occurring when a neuronal membrane potential crosses threshold. Each channel (up to 4 per electrode, scanned at 40 KHz) is associated with a specific neuron via its action potential shape (spike sorting).  Cell-electrode coupling is stable so that wave-shapes can be followed with accuracy, for many channels and over many days at a time. 

The results presented below were obtained from five hour recordings taken from three different cortical cultures, spanning a range of sizes, with 20,  33 and 62 recorded cells. The three networks are mature (see e.g. \cite{Tateno_2002}), with ages {\it in vitro} of 34, 51 and 42 days, respectively. The number of cells recorded are a fraction (estimated roughly at 5-10\%) of all active neurons in the culture.
The total activity rates for the network were stable over the course of the recordings. To construct binary states (zero/one) of multiple time series we sampled them at one half the spike rate of the most active cell. This gives the highest entropy per measured bit for a temporally uncorrelated data stream.

\section{Information theoretical analysis of neuronal spike trains}
\label{sec:info-theory}

Analyzing spike trains to understand how groups of neurons share and process information is challenging because neuronal time series typically display strong temporal stochasticity.  This is the result of both the spatial sparseness of the collected signals of individual neurons, which receive and send processes to many other hidden (not recorded) cells, and of the intrinsically non-linear and stochastic  nature of neuronal and synaptic dynamics \cite{BookNeuroscience}. 

\begin{figure}
\begin{minipage}[b]{0.45\linewidth}
 \includegraphics[angle=0,width=3.4in]{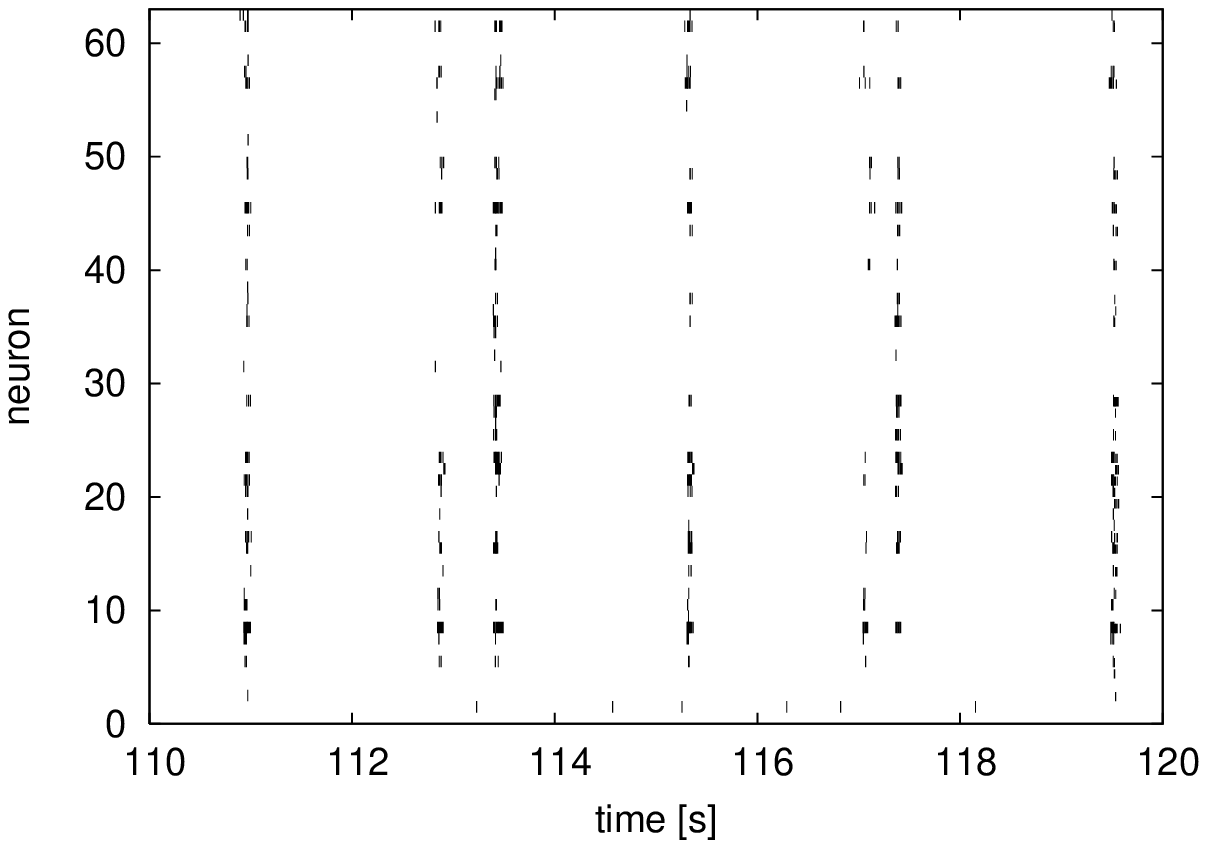}
\end{minipage}
\hspace{1.2cm}
\begin{minipage}[b]{0.45\linewidth}
\includegraphics[angle=0,width=3.4in]{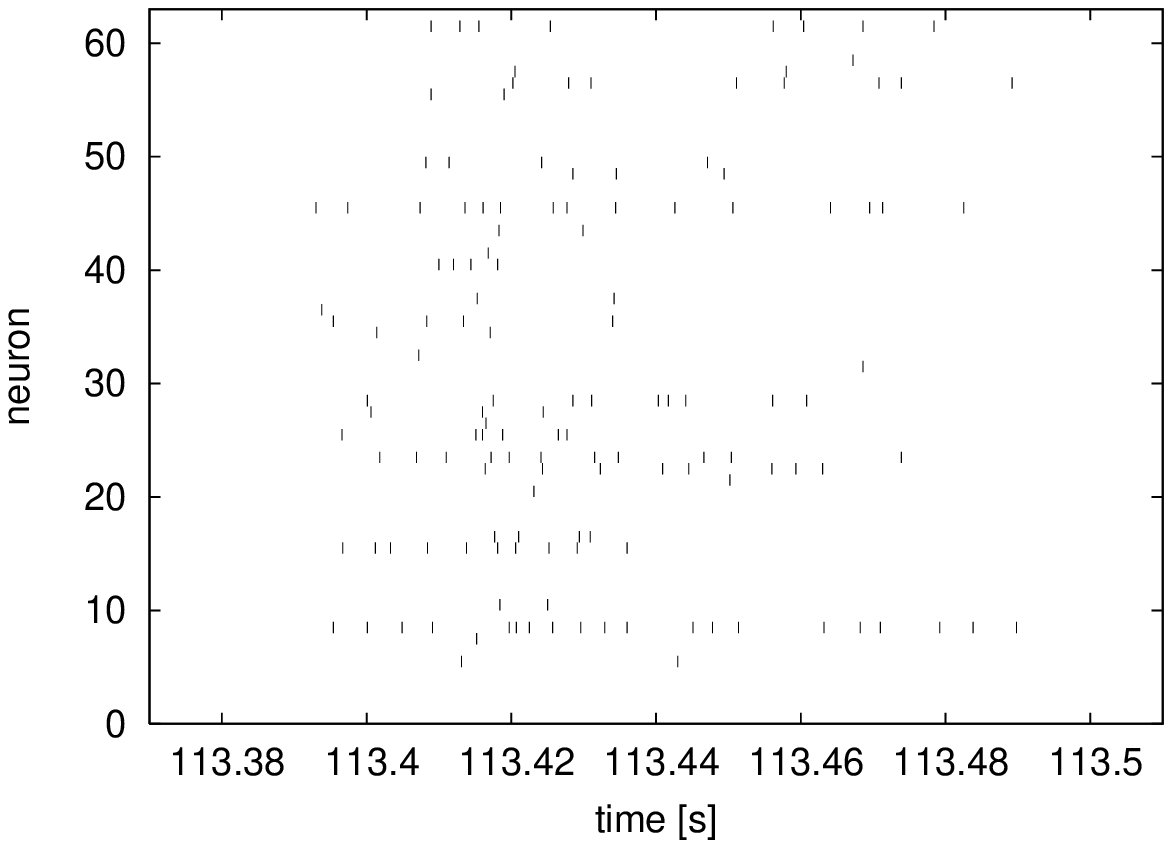}
\end{minipage}
\caption{Spike time series for the spontaneous (native) activity of a cortical culture with 62 recorded neurons.  Much of the activity takes place as part of network-wide collective events (left, vertical features),  known as network bursts.  The structure of each neuron's spike train observed over short time scales (right, network burst detail) shows strong stochasticity.}
 \label{fig:spiketrains}
\end{figure}

The noisy character of neuronal spike trains, see Fig.~\ref{fig:spiketrains},  suggests a probabilistic description of their time series,  in terms of sequences of active spiking periods and silences \cite{Spikes}, as strings of zeros and ones. This standard decomposition allows the analysis of neuronal time series as noisy channels via information theory. In particular  we can measure how the activity of one cell is informed by the state of others and determine their relative configuration, which will allow us to reconstruct proxy connectivity networks mapping how information is shared, stored and processed collectively in these ensembles. 

\subsection{Mutual information between neurons}
\label{ssec:mut-info}

Mutual information provides a natural quantitative measure of the degree of statistical dependence between two stochastic variables, free of assumptions about the nature of the underlying joint distribution.  As such, mutual information is a non-linear generalization of familiar correlation functions \cite{Li}.  In the neuronal networks under consideration here, large mutual information between suitably defined time series of two neurons will be taken as an indication of connectivity between them.  Subtleties that  arise with this association will also be discussed and, at least partially, resolved below.

We start by recalling the definition of the mutual information between two stochastic variables $X$,$Y$
\begin{eqnarray} 
I(X;Y) = \sum p(x,y) \log_2 \left( \frac{p(x,y)}{p(x) p(y)} \right), 
\label{mutInfo}
\end{eqnarray}
where $p(x,y)$ is the joint distribution and $p(x) , p(y) $ are the single variable marginals.  The mutual information is semi-positive definite, and equal to zero if and only if the two variables $X, Y$ are statistically independent, i.e. $p(x,y)=p(x)p(y)$.  

The magnitude of the mutual information between two cells is dependent on the information content of the time series of each cell, in terms of their Shannon entropy $H(X)$, $H(Y)$. This becomes clear through consideration of the relations
\begin{eqnarray}
I(X;Y) = H(X)-H(X|Y)=H(Y)-H(Y|X) \quad \rightarrow  I(X;Y)\leq {\rm min}\left[ H(X),H(Y) \right].
\end{eqnarray}
This bound, taken together with $I(X;Y)\geq 0$, implies that the normalized mutual information $i(X;Y)$
\begin{eqnarray}
i(X;Y) = \frac{I(X;Y)}{{\rm min}\left[ H(X),H(Y) \right]}
\end{eqnarray}
is always in the range  $0\leq i(X;Y)\leq 1$, and can be used to measure the strength of the link between two cells.  In this spirit, more restricted measures of statistical dependence such as cross-correlation
have also been used to identify putative links between simultaneously recorded neurons \cite{CrossCorr}. 

\subsection{Multiple cell functional arrangements}
\label{ssec:R-info}

\begin{figure}
 \begin{minipage}[b]{0.9\linewidth}
 \includegraphics[angle=0,width=4.5in]{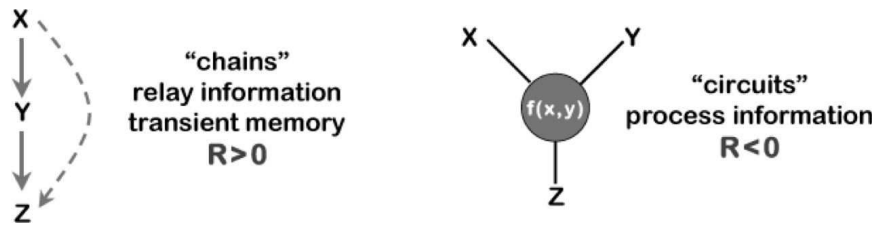}
 \end{minipage}
\caption{Different connectivity arrangements of neurons have distinct roles in short term memory or in information processing. These  can be identified via the information theoretical analysis of multi-cell spike time series, with the former associated with $R>0$ and the latter $R<0$, see text.} 
 \label{fig:funcchain}
\end{figure}

We would like to not only reconstruct a network reflecting putative connectivity between neurons, but also to identify how information is relayed or processed through neuronal activity. Fig.~\ref{fig:funcchain} distinguishes relaying and processing computations  for arrangements of three neurons. Sequential chains of cells can relay information, and are known to play a part in short term memory in the brain \cite{ChainsMemory1,ChainsMemory2}.  On the other hand neurons receive inputs from many other cells, so that their spiking activity may behave as the (non-linear) function of multiple inputs. In this way individual neurons naturally integrate activity from many sources and are able to process information. Here we show how these two different types of structures can be distinguished given data streams for multiple cells. Consequently we can determine which type of functional structure is more prevalent in cortical cultures.   While three-cell networks are still extremely small relative to even a single cortical column 
(approximately $10^5$ neurons),  they are complex enough to illustrate the elementary properties of network computation.

The computations performed by networks of neurons can be analyzed via the consideration of appropriate multi-informations between groups of cells.  To see this consider that the notion of mutual information, Eq.~(\ref{mutInfo}), naturally generalizes to higher dimensions. For example $I(X_1,X_2,\ldots,X_n)$ is the mutual information between the joint distribution of $n$ variables $p(\{x_i\})$ and the product of each single variable's distribution. This quantity is given explicitly for n=3, for several examples in table~\ref{TableI}\ref{tabI} . Similarly we will write  $I(\{X_1,X_2,\ldots,X_n\};Y)$ to denote the mutual information between the joint state of $n$ variables $\{X_i\}$ and variable Y.

The use of  multi-information between groups of variables is useful because it allows us to examine connectivity arrangements involving more than just pairs of elements.  In particular, for an arrangement of three elements, we consider the conditional mutual information of two of the variables $X, Y$, subject to the state of a third $Z$  \cite{CoverThomas}, 
\begin{eqnarray}
I(X;Y|Z) =  \sum_{x,y,z} p(x,y,z) \log_2 \frac{p(x,y|z)}{p(x|z) p(y|z)}.  
\label{defCondI}
\end{eqnarray}
$I(X;Y|Z)$ is a mutual information  and as such is semi-positive definite and zero only when $X, Y$ are independent stochastic variables (each conditioned on $Z$). Thus $I(X;Y|Z)$ tells us whether consideration of a third variable supplies more or less information about the nature of the correlation between $X$ and $Y$.

This property suggests the consideration of the difference
\begin{eqnarray}
R(X,Y,Z)= I(X;Y) - I(X;Y|Z)
\label{defR1}
\end{eqnarray}
as a compact means to characterize how knowledge of $Z$ informs us of the state of $\{X, Y\}$.
Although perhaps not immediately obvious this quantity is identical to that recently proposed by Schneidman {\it et al.} 
\cite{Schneidman2003,Schneidman_JNeuroscience_2003_synergy} as a measure of redundancy of the knowledge of $Z$ given $X,Y$. In their definition, for three variables
\begin{eqnarray}
R(X,Y,Z)= I(X;Z) +I(Y;Z) - I(\{X,Y\};Z).
\label{defR2}   
\end{eqnarray}
We prove in Appendix A that the two definitions are equivalent, where we also show that $R$ is fully symmetric under permutations of its arguments, and consequently can be computed more parsimoniously than its definition may at first suggest.

It is immediately clear from the definition of $R$ that it can be positive, negative or zero. A positive $R$ is equivalent to $I(X;Y)>I(X;Y|Z)$  indicating that knowledge of $Z$ does not improve (and may degrade) knowledge  of $\{X, Y\} $ correlations. Conversely $R<0$ indicates that  knowledge of $Z$  leads to more complete information about $\{X, Y\}$. Finally $R=0$ indicates that at least one of the three variables, is independent of the others. In this sense $R=0$ denotes the absence of a  genuine three variable structure. After Schneidman {\it et al.} \cite{Schneidman2003} we refer to these three situations as redundancy ($R>0$), synergy ($R<0$) and independence ($R=0$).  To better convey their meaning we illustrate each of these cases with examples. A summary of what can be learned from these quantities about the detailed connectivity mapping between variables is given in Table~\ref{InfoConnectivity}.

\subsubsection{Independence: $R=0$}
\label{sssec:R=0}

It follows immediately from definition (\ref{defCondI}) that if $Z$ is independent of $\{X,Y\}$ then $R=0$. Indeed if both $X$ and $Y$ are statistically independent of $Z$, so that 
$p(x|z)=p(x)$, $p(y|z)=p(y)$ and naturally $p(x,y|z)=p(x,y)$. Then $I(x,y|z)=I(x,y)$, which implies $R=0$. 

Moreover this is true in general, as expected from symmetry of $R$, if any of the three variables is independent of the other two. We see this for the Boolean  functions $X$ and $Y$, of tables \ref{tabBF} and \ref{tabI}.   In these cases we would infer the existence of only one link from the only non-zero binary mutual information, and the absence of a three variable structure, from $R=0$.

\subsubsection{Synergy: $R<0$}
\label{sssec:R<0}

If $Z$ is a function of {\em both} variables $X,Y$ then knowledge of $Z$ leads to greater information of the joint state of $\{X,Y\}$. In this case $R<0$, and we speak of {\it synergy}.

In order to illustrate this point we follow Schneidman  {\it et al.} \cite{Schneidman2003} in considering a set of Boolean  functions of two binary variables $X, Y$. Although these functions are much too stylized to be realized between neurons, they supply us with the simplest useful illustrations. Table ~\ref{tabBF} shows several of the sixteen possible Boolean functions of two binary variables.

\begin{table}[htdp]
\subtable[Examples of the relationships among three Boolean variables, showing synergy, independence or redundancy, see text. Nontrivial logical circuits (e.g. OR, AND) are examples of functions for which output is a joint function of the two inputs and are thus associated with information processing. Outputs that reflect only one of the inputs are trivial as three cell structures (independence). CHAIN is an example of a third type of relation where variables are not independent, but knowledge of any third cell does not convey new information.]{
\begin{tabular}{c c c c c c c c c c c}
\hline
\multicolumn{2}{c}{Inputs} &  \multicolumn{5}{c}{Synergy}  & \multicolumn{2}{c}{Independence} &  \multicolumn{2}{c}{Redundancy} \\ \hline  
x & y  &\quad   AND & OR & XOR  & Y$\dashv$X & X$\dashv$Y & \qquad X   & Y & \multicolumn{2}{c}{CHAINS}  \\ \hline 
0 & 0 &  \quad  0       &  0    &  0      & 0         & 0  & \qquad 0  & 0  & \quad 0  &  1\\ 
0 & 1 & \quad  0        &  1    &  1      & 0         & 1  &  \qquad 0  & 1 &  \quad - &   - \\
1 & 0 & \quad  0        &  1    &  1      & 1         & 0  &  \qquad 1  & 0  & \quad -  &   - \\
1 & 1 & \quad  1        &  1    &  0      & 0         & 0  &  \qquad 1  & 1 & \quad 1 &    0\\
\hline
\end{tabular}
\label{tabBF}
}
\qquad\qquad
\subtable[Information theoretic quantities for the Boolean functions of Table~\ref{tabBF}. Inputs for $X$, $Y$ are taken at random so that $H(X)=H(Y)=1$, except for CHAIN, where x=y, resulting in the only case where there is mutual information between the two inputs. The functions Y$\dashv$X and X$\dashv$Y give the same results as AND and were thus omitted for simplicity.  Note that there are possible connectivity structures, like XOR, which have zero mutual information between any two links, but still exist as synergetic triplets.]{
\begin{tabular}{c c c c c c c c }
\hline
 Function   & H(Z)   			    & I(X;Y) & I(X;Z) & I(Y;Z) & I(X,Y,Z) & R   \\ \hline 
AND     & $2-\frac{3}{4}\ln_2 (3)$  &  0   & $ \frac{3}{2} -\frac{3}{4}\ln_2 (3)$  & I(X;Z) & $2-\frac{3}{4}\ln_2 (3)$  &$1- \frac{3}{4}\ln_2 (3)$\\ 
OR         & $2-\frac{3}{4}\ln_2 (3)$  &  0   & $ \frac{3}{2} -\frac{3}{4}\ln_2 (3)$  & I(X;Z) & $2-\frac{3}{4}\ln_2 (3)$  &$1-\frac{3}{4}\ln_2 (3)$ \\
XOR      & 1  				     &  0  &  0 & 0 & 1&  -1   \\ \hline
X            & 1  				     &  0  &  1 & 0  & 1 & 0  \\
Y            & 1  				    & 0  &  0 & 1 & 1 &  0  \\ \hline
CHAIN & 1 				    & 1  &  1&  1 & 2 & 1  \\ \hline
\end{tabular}
\label{tabI}
}
\caption{Examples of 3 cell functional arrangements and their information theoretical signatures.}
\label{TableI}
\end{table}

In all the cases of Table~\ref{tabBF}, except for CHAIN to be discussed below,  $X$ and $Y$  are independent  random variables ($p_0=p_1=1/2$) and are thus statistically independent so that $I(X;Y)=0$. Information theoretic quantities for these functions are given in Table~\ref{tabI}.

Considering exclusively the mutual information in Table~\ref{tabI} as a starting point we see that we would infer correctly that $X, Z$ and $Y, Z$ are connected but $X, Y$ are not for the functions AND, OR (and $X \dashv Y$ and $Y \dashv X$, not shown). 

We may find it surprising however that for XOR we would infer no connectivity (all pairwise mutual informations are zero). We find nevertheless a negative $R$ and a nonzero $I(X,Y,Z)$, which tell us that  the three variable joint distribution is non-trivial. Thus we infer a genuine three cell object present, despite the absence of pairwise mutual informations. Such structures would be missed under exclusively pairwise analysis.
 
Finally we notice that a fully connected diagram, in which variables are not Markovian (see below) is also synergetic and therefore implies $R<0$.  
 
\subsubsection{Redundancy: $R>0$}
\label{sssec:R>0}

Finally we consider cases where knowledge of a third variable $Z$ provides redundant information about to the state of $\{ X,Y\}$. The simplest examples are the functions CHAIN of Table \ref{tabBF}. 
CHAIN can be seen as the limit of the (anti)ferromagnetic interaction considered in Schneidman {\it et al.} \cite{Schneidman2003}, as the coupling $\beta \rightarrow \infty$. 

Clearly the state of the third variable, while not independent, does not add information about the state of the other two.  This situation is generic of connections  organized as chains. An important and tractable situation is when variables are Markovian, in which case a well known result, the {\it information processing inequality}, emerges. This implies in turn that  $R>0$, generally, so it is worth explicit consideration.

If $X$, $Y$, $Z$ form a Markov chain, denoted $X\rightarrow Y \rightarrow Z$, then their joint distribution can be written as \cite{CoverThomas} 
\begin{eqnarray}
p(x,y,z)=p(x)p(y|x)p(z|y).
\end{eqnarray}
This means in particular that $X$, $Z$ are independent given $Y$, i.e.  $I(X;Z|Y)=0$.
In this situation the information processing inequality \cite{CoverThomas} also tells tells us that for a Markov chain $X\rightarrow Y \rightarrow Z$
\begin{eqnarray}
I(X;Y) \geq I(X;Z).
\end{eqnarray}
Now, writing $I(\{X,Y\};Z)$ as (see Appendix A) 
\begin{eqnarray}
I(\{X,Y\};Z) = I(X;Z)+I(X;Y|Z) = I(X;Y)+I(X;Z|Y)
\end{eqnarray}
and using $I(X;Z|Y)=0$ we conclude that 
\begin{eqnarray}
R=I(X;Y) - I(X;Y|Z)=I(X;Z) \geq 0.
\end{eqnarray} 
The same result is obtained for the chain $Z\rightarrow Y \rightarrow X$, which from the point of view of information quantities is equivalent to its reversed order considered above.

For the chain $X\rightarrow Z \rightarrow Y$, for which $I(X;Y|Z)=0$, it follows immediately that $R=I(X;Y)\geq 0$. In particular if the variables are indeed connected then $I(X;Y)>0$ strictly  and $R >0$ is a general property of Markov chains.

Finally we note that if $Z$ drives both $X$ and $Y$, as in $X \leftarrow Z \rightarrow Y$ then $R>0$. 
We may suspect this from the fact that $Z$ is then a function of a single variable. Nevertheless formally  this follows from the form of the joint distribution  
\begin{eqnarray}
p(x,y,z) = p(z) p(x|z) p(y|z) \Rightarrow p(x,y|z)=\frac{p(x,y,z)}{p(z)}= p(x|z) p(y|z), 
\end{eqnarray}
which implies $I(X;Y|Z)=0$\, and therefore that $R=I(X;Y) > 0$. 
In the case of $R>0$ we therefore preserve only the two strongest links among the three cells, a strategy also adopted in \cite{Margolis_2006}.

\subsubsection{Normalization of R}
\label{sssec:R-norm}

Just as for the mutual information the absolute value of $R$ is dependent on the magnitude of the various multi-informations involved in its definition. We can place upper and lower bounds on $R$, so that a normalized quantity $r(X,Y,Z)$ can be constructed, that factors out these effects.  

First note that it follows directly from the definitions (see also appendix A) that 
\begin{eqnarray}
R \leq {\rm min} \left[ I(X;Y), I(X;Z); I(Y;Z) \right].
\end{eqnarray}
So that if $R>0$, $r=R/{\rm min} \left[ I((X;Y), I(X;Z); I(Y;Z) \right]$. Similarly when $R<0$ then 
\begin{eqnarray}
R \geq - {\rm min} \left[ I(X;Y|Z), I(X;Z|Y), I(Y;Z|X) \right].
\end{eqnarray} 
Note also that 
\begin{eqnarray}
R \geq - {\rm min} \left[ I(\{X,Y\};Z), I(\{X,Z\};Y), I(\{Y,Z\};X) \right],
\end{eqnarray} 
but that the first bound is more stringent (see appendix A). Thus for $R<0$ a normalized $r$ is given by
\begin{eqnarray}
r=\frac{R}{{\rm min} \left[ I(X;Y|Z), I(X;Z|Y), I(Y;Z|X) \right]}.
\end{eqnarray} 

\section{The structure of networks of spiking neurons {\it in vitro}}
\label{sec:network}

In this section we show how to use the information theoretic diagnostics of the previous section 
to construct connectivity graphs for neuronal ensembles.

\subsection{Graph construction in practice}
\label{ssec:graph-construction}

The joint consideration of $R$ and of pairwise mutual informations leads to a classification of the connectivity diagrams of any three nodes. The summary of these results, which can be implemented as an algorithm to map the network connectivity, is summarized in Table~\ref{InfoConnectivity}. An efficient algorithm computes first the value of $R$ for a chosen triplet of variables. This determines how many links to attribute.  The strength of  these links, whenever appropriate, are determined by mutual information between pairs of variables. Whenever the consideration of several trios implies different weights for the same connection between two neurons, the results are averaged.

To determine whether measured values of $R$ and $I$ are significant, we compute their values according to a null model where every cell $i$ in the network produces spikes at the same average rate $\lambda_i$ as observed for each neuron's actual spike train, but whose temporal structure is otherwise random.  This corresponds to a number of spikes per unit time given by a Poisson distribution with rate $\lambda_i$ or equivalently, as we do in practice, a time interval between consecutive spikes given by an exponential distribution.   Under perfect estimation this Poisson model generates statistically independent spike trains between any two cells. In practice estimation is imperfect and  this procedure leads instead to very small, but non-zero, values of $R$ and $I$. Thus for each trio of cells we take the observed $R$ to be significant if it is larger than three times that measured via the Poisson model.  Otherwise the trio is considered statistically independent with $R=0$. We proceed similarly for the mutual information measured for any two cells.

\begin{table}[htdp]
\begin{tabular}{l | l l l l}
\hline
\multirow{3}*{$R=0$} & \multicolumn{4}{c}{No 3-cell connected structure} \\ & & \multicolumn{3}{c}{\qquad} \\
& No connections: & \multicolumn{3}{c}{$\forall_{i,j \in \{1,2,3\}} I(X_i,X_j)=0$} \\ & & \multicolumn{3}{c}{\qquad} \\
& 1 connection: & \multicolumn{3}{c}{$\exists !_{i,j \in \{1,2,3\}} I(X_i,X_j) \neq 0$} \\
\hline 
\multirow{4}*{$R \neq 0$} & \multicolumn{4}{c}{There is a 3-cell connected structure} \\ & & \multicolumn{3}{c}{\qquad} \\
& 2 connections: & \multicolumn{3}{c}{$R>0$ (redundancy)}   \\
& \qquad & $X \leftrightarrow Y \leftrightarrow Z$ & $X \leftrightarrow Z \leftrightarrow Y$ & $X \leftarrow Z \rightarrow Y$ \\ 
& \qquad & Markov: & Markov: & \qquad  \\
& \qquad & $I(X;Y)\geq I(X;Z)$ & $I(X;Z) \geq I(X;Y)$ & \qquad \\
&\qquad & $I(X;Z|Y)=0$ & $I(X;Y|Z)=0$ & $I(X;Y|Z)=0$ \\
& \qquad & $R=I(X;Z)$ & $R=I(X;Y)$ & $R=I(X;Y)$ \\ & & \multicolumn{3}{c}{\qquad} \\
& 3 connections: & \multicolumn{3}{c}{$R<0$ (synergy)} \\
& \qquad &  \multicolumn{3}{c}{$X\rightarrow Z \leftarrow Y$: $Z=f(X,Y)$} \\ 
& & \multicolumn{3}{c}{\qquad Each variable is a function of all others.} \\
\hline
\end{tabular}
\caption{The joint consideration of the values of  $R$ and of the binary mutual informations $I$ lead to a classification of the connectivity arrangements between three stochastic channels. For $R>0$ only the two strongest links are adopted. For $R<0$ we adopt three links as a general representation of the specific functional interdependence between the three cells.} 
\label{InfoConnectivity}
\end{table}

\subsection{Graph Theoretic Analysis}
\label{ssec:graph-analysis}

\begin{figure}
 \begin{minipage}[b]{0.9\linewidth}
 \includegraphics[angle=0,width=4in]{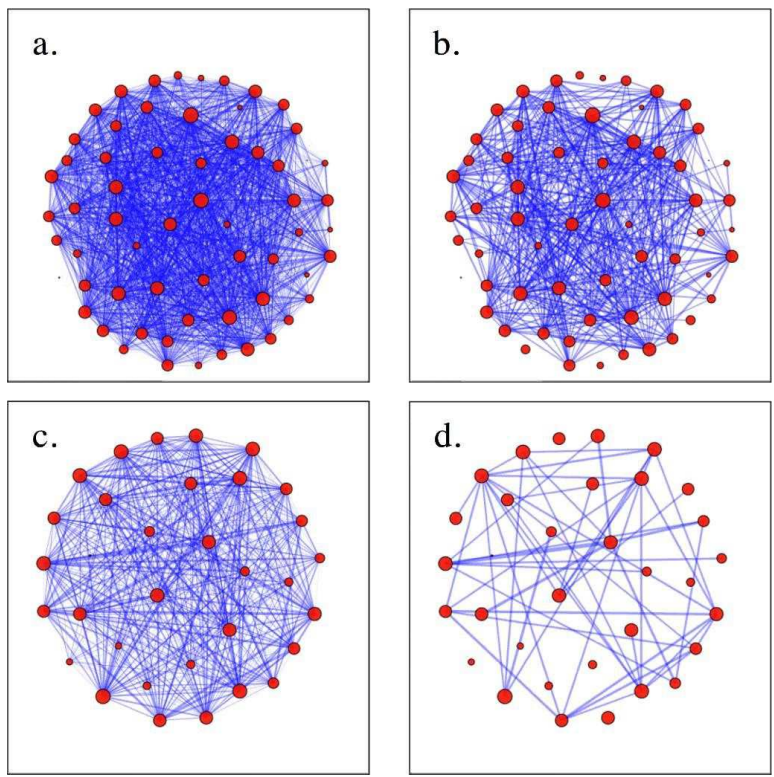}
 \end{minipage}
\caption{Weighted graphs for network of a) 62 and c) 33 recorded neurons. Panels b) and d) show the corresponding networks selecting only the links with strength above 0.4.  The graphs are characterized by a large number of weak links that connect almost every pair of cells, and of fewer stronger links that tend to interconnect subsets of neurons. Node size is drawn proportionally to its total connectivity weigh.} 
 \label{fig:networks}
\end{figure}

\begin{figure}
\begin{minipage}[b]{0.4\linewidth}
\includegraphics[angle=270,width=3in]{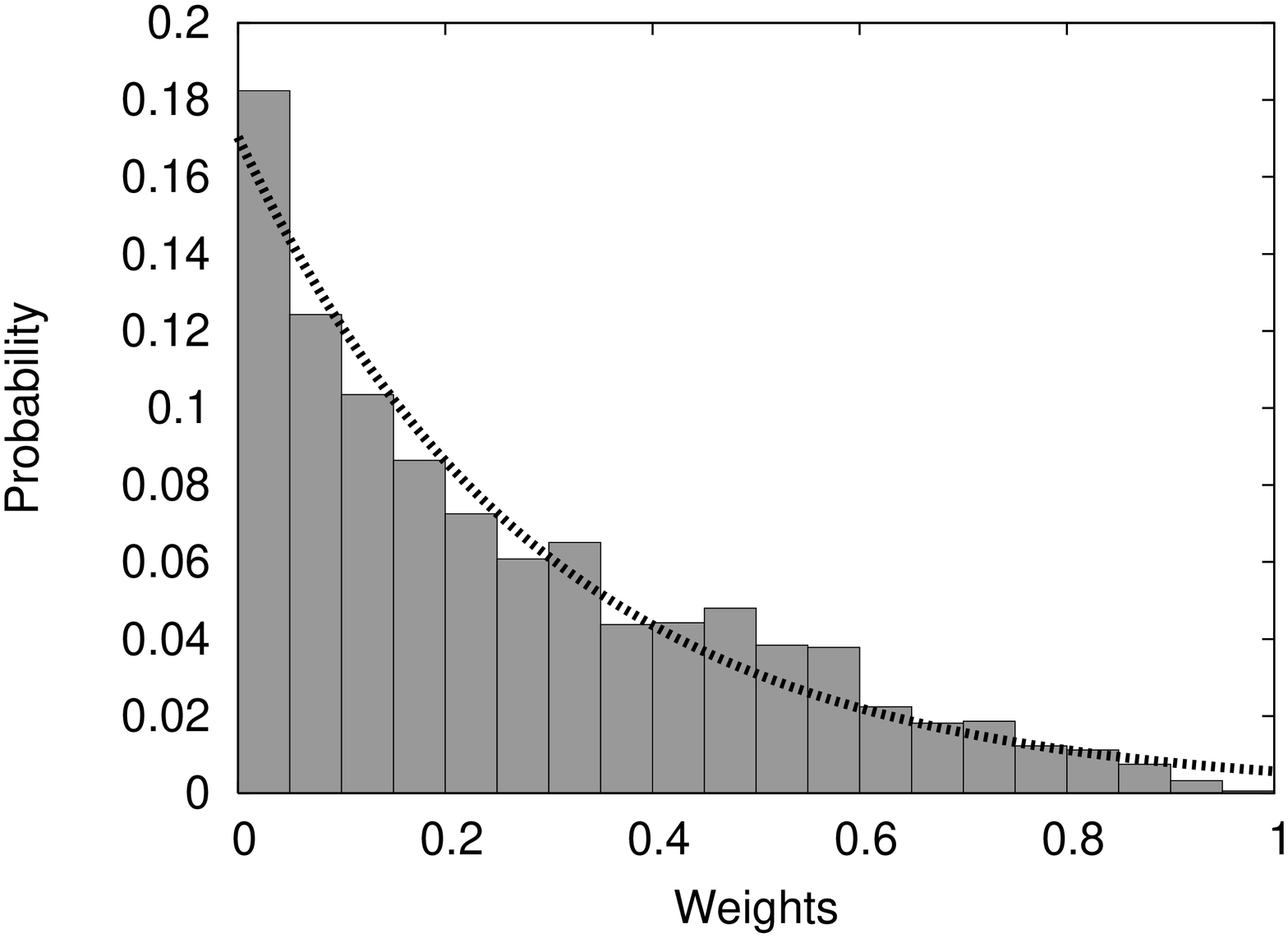}
\end{minipage}
\hspace{1.2cm}
\begin{minipage}[b]{0.4\linewidth}
 \includegraphics[angle=270,width=3in]{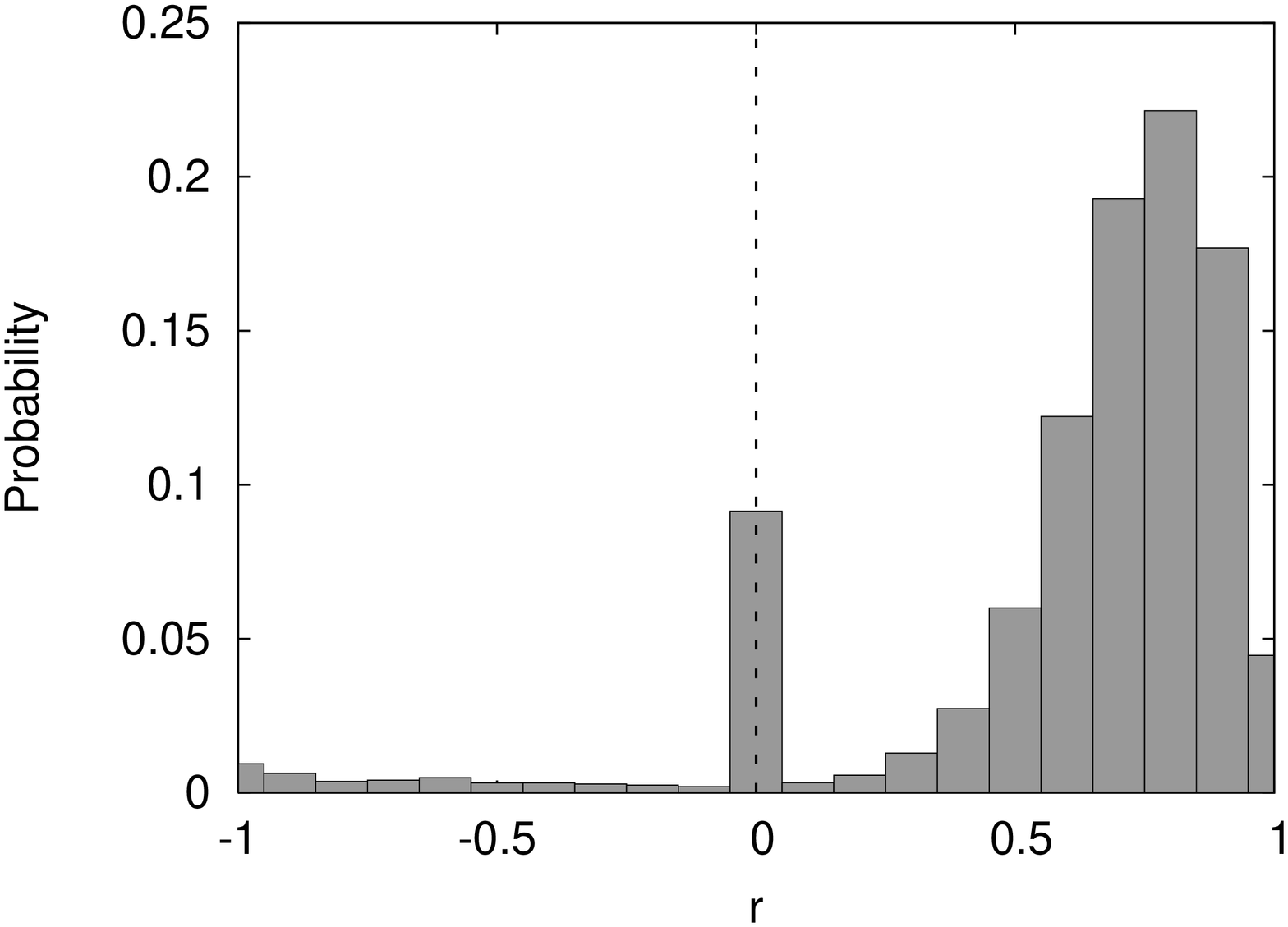}
\end{minipage} 
\caption{The probability distribution for weights of the inferred network of 62 recorded cells (left). This distribution is well described by an exponential (dashed line), with average connectivity weight $0.294\pm 0.019$. The distribution of normalized $r$ for the same network (right). Most trios show information redundancy ($r>0$), while a smaller number display synergy ($r<0$). The central peak at $r=0$, with probability $p=0.091$, consists of a fraction of $70\%$ of truly independent trios ($R=0$), $22\%$ with negative $r \geq -0.05$ and $8\%$ with positive $r \leq 0.05$.
The other two networks exhibit quantitatively consistent results, but with poorer statistics. These results are suggestive that multi neuron connectivity structures associated with both information processing ($R<0$) and short term memory ($R>0$) are present in cultured neuronal networks {\it in vitro}. } 
\label{fig:connectivity-dist}
\end{figure}

Next we characterize the structural properties of the resulting networks in terms of graph theoretic quantities. Although all three graphs are small in terms of numbers of nodes we show that results about their network structure are consistent for the different networks. Figure~\ref{fig:networks} shows the weighted graphs for two of these networks. Figure~\ref{fig:connectivity-dist} shows the connectivity weight distribution and the histogram of $r$ for all cell trios.  

The distribution of connectivity weights is well-described by an exponential. This implies that most links are weak, with a few stronger ones tending to inter-connect subsets of cells. The distribution of $r$ shows that most trios are associated with redundant structures that may be involved in short term memory functions. Nevertheless a  number of trios display $r<0$, indicating the existence 
of non-trivial functional relations between measured cells, that may play a role in information processing. 

To compare the structure of these weighted graphs quantitatively to other complex networks we transform them into binary graphs, with each link set to zero or one if its strength lies below or above a given threshold, respectively. Table~\ref{table:net-diag} shows the average clustering coefficient, diameter and Pearson coefficient for degree-degree correlations for the three networks as functions of the threshold, and compares them to results from one hundred realizations of the standard null model, {\it viz}. a network with the same node degrees but with randomized connections.   

We see that in every case the neuronal graphs display significantly higher clustering than the null networks. They also show small diameters of the largest component $d$, of order $d\sim \ln(N)$, where $N$ is the number of nodes. These two properties together indicate that cortical networks {\it in vitro}, like the neuronal network of the nematode  \cite{Watts}, are {\it small world graphs}. 

Finally we observe that cortical networks   {\it in vitro} are weakly disassortative, i.e. that the 
Pearson coefficient associated with degree-degree correlations is, in most cases, negative but small    \cite{Newman}. This indicates that nodes of dissimilar degree preferentially connect to each other. 
Together these measures show that networks of neurons, constructed via the information shared by groups of cells,  are significantly different from randomized graphs, but share instead many structural properties with other complex networks \cite{Watts,BookNetworks1,BookNetworks2}.

\begin{table}[htdp]
\begin{center}

\begin{tabular}{c c c c c c c }
\hline
 Size & Threshold   & \multicolumn{2}{c}{Av. Clustering} & \multicolumn{2}{c }{Diameter} & Assortativity \\
 (recorded neurons N)             & $[0,1]$         & Actual & Random & Actual & Random  & Pearson coeff. \\ \hline
\multirow{3}*{ N=20 } & 0.1 & 0.89 & 0.52$\pm$ 0.04 & 2 & 2.97$\pm$ 0.03 & -0.150 $\pm$ 0.025 \\
 & 0.3 & 0.57 & 0.24 $\pm$ 0.06 & 4 & 4.15 $\pm$ 0.55 & -0.040 $\pm$ 0.048  \\  
 & 0.5  &  0.32 & 0.05 $\pm$ 0.10 & 1 & 2.10 $\pm$ 0.70  & -0.200 $\pm$ 0.100 \\ \hline
\multirow{3}*{ N=33 } & 0.1 & 0.83 & 0.48$\pm$ 0.02 & 3 & 2.97$\pm$ 0.03 & -0.114 $\pm$ 0.001\\
 & 0.3 & 0.61 & 0.26 $\pm$ 0.04 & 4 & 3.99 $\pm$ 0.30 &  -0.033 $\pm$ 0.026   \\ 
 & 0.5  &  0.07 & 0.00 $\pm$ 0.02 & 2 & 2.98 $\pm$ 0.86  & 0.064 $\pm$ 0.053 \\ \hline
\multirow{3}*{N=62} & 0.1 & 0.82 & 0.51$\pm$ 0.01 & 3 & 2.97$\pm$ 0.03 & -0.025 $\pm$ 0.004  \\
 & 0.3 & 0.60 & 0.37 $\pm$ 0.02 & 4 & 4.00 $\pm$ 0.20 & -0.093 $\pm$ 0.004  \\ 
 & 0.5  &  0.43 & 0.24 $\pm$ 0.02 & 3 & 3.93 $\pm$ 0.18  & -0.102 $\pm$ 0.006 \\ \hline 
 \end{tabular}
\end{center}
\caption{Graph theoretic measures for each of three cortical neuronal networks and as  functions of the  weight threshold, and their comparisons with the average and standard deviation obtained from a graph with the same degree but randomized connections. 
Cultured neuronal networks are clearly distinguishable from randomized networks with the same degree. They are {\it small world graphs} with weakly disassortative degree correlations.}
\label{table:net-diag}
\end{table}

\section{Discussion and Conclusions}
\label{sec:conclusions}

The understanding of the functional network structures used by living nervous systems to store and process information is key to unravelling the mechanisms of their computational power. It is now possible to measure the activity of networks of individual neurons and to start to identify such informational structures, their development over time and their role in response to external stimuli (see e.g \cite{Schneidman_Ising}).
Here we have shown how time series for networks of spiking neurons can be treated statistically and analyzed in terms of informational theoretic quantities to reveal how information is shared, relayed and processed by groups of cells.  This analysis generates networks of shared information, which are proxies for the physical connectivity between neurons. In fact, knowledge of this type of connectivity and its dynamics may well suffice to reveal the underpinnings of  collective computation in networks of neurons. 

We have also shown that consideration of information theoretic quantities beyond two cells can identify structures that are naturally higher order, revealing for example a cell whose output is a non-linear function of several others. The specificity of such structures is lost when we represent them as sets of binary links, as we have done here to produce standard network analyses. The systematic mapping and characterization of these functional modules goes beyond the scope of the present work, and will be pursued elsewhere.

Neglecting such functional specificity we have built networks of shared information, which we showed to be weakly disassortative {\it small world graphs}, in agreement with e.g. the detailed neuronal network of {\it C. elegans} \cite{Watts}, but in contrast to networks of correlation between large scale brain regions, identified via fMRI measurements \cite{fMRI}.
The distribution of weights is approximately exponential and multi neuron  structures associated with information relaying dominate over a smaller number of cell arrangements associated with information processing. Regardless of details of specific proportions for these cellular arrangements we showed, by identifying multi-neuron structures with $R>0$ and $R<0$,  that neuronal modules associated with short term memory and information processing co-exist in disassociate neuronal cultures {\it in vitro}. How these structures and their generalizations to larger cell numbers may be harnessed to perform reliable computation remains perhaps the clearest single challenge for future studies of information processing in  neuronal networks living {\it in vitro}.

These results raise several other interesting  questions for future research. It is natural to expand the analysis presented here to include multi-information quantities involving more cells, in order to identify the functional role of more complex neuronal arrangements. It is also interesting to investigate their change over time, and under electrophysiological stimulation. Finally, this type of analysis can be applied to any distributed information processing system, natural or artificial, and should prove useful in revealing the computational structures that constitute other types of nervous tissue, {\it in vivo} and {\it in vitro}, and in creating mathematical models and engineered systems that mimic them.

\section*{Acknowledgments}

This work was carried out in part under the auspices of the National Nuclear Security Administration of the U.S. Department of Energy at Los Alamos National Laboratory under Contract No. DE-AC52-06NA25396. Additional support from LDRD-ER project 20050411.  LMAB thanks Jim Crutchfield, Aric Hagberg and Pieter Swart for useful discussions.

\section*{Appendix A }

In this appendix we prove the equivalence of the definitions of $R$, Eqs.~(\ref{defR1}) and (\ref{defR2}). This equality establishes that the measure $R$ of redundancy, proposed by Schneidman {\it et al.} in Ref. \cite{Schneidman2003}, can be written, for three variables in terms of the difference between the mutual information of a pair of variables and its mutual information conditional on a third cell (\ref{defR1}).

We begin with the definition of $R$ given two variables X, Y, and upon consideration of a third Z, which is a candidate for having connections with X, Y or may be an external stimulus
\begin{eqnarray}
R = I(X;Z) +I(Y;Z) - I(\{ X,Y \};Z),
\label{def_R}
\end{eqnarray}
where, as usual, the semicolon separates the states being compared, so that the last mutual information is between the joint state of $\{ X, Y \}$ and Z. 
We will now prove that $R$, defined in (\ref{def_R}) can also be written as  
\begin{eqnarray}
R= I(X;Y) - I(X;Y| Z).
\end{eqnarray}

Starting from (\ref{def_R})  we use the chain rule for mutual information \cite{CoverThomas}
\begin{eqnarray}
I(\{X_1,X_2, \dots, X_n\};Z)= \sum_{i=1}^n I(X_i;Z|X_{i-1}X_{i-2}, \ldots, X_1).
\label{Iexpansion}
\end{eqnarray}
Specifically for three variables (n=2) 
\begin{eqnarray}
&& I(\{X,Y\};Z)= I(X;Z) + I(Y;Z|X) = I(Y;Z)+I(X;Z|Y),  \\
&& I(\{Y,Z\};X)= I(X;Y) + I(X;Z|Y) = I(X;Z)+I(X;Y|Z),  \\
&& I(\{X,Z\};Y)= I(Y;Z) + I(X;Y|Z) = I(X;Y)+I(Y;Z|X).  
\end{eqnarray}
These expression can now be  used in (\ref{def_R}) to give
\begin{eqnarray}
I(X;Z) +I(Y;Z) - I(\{X,Y\};Z) = I(Y;Z) - I(Y;Z|X)=I(X;Z)-I(X;Z|Y).
\end{eqnarray}
Similarly we can write 
\begin{eqnarray}
 I(X;Y) +I(Z;Y) - I(\{X,Z\};Y) = I(X;Y) - I(X;Y|Z)=I(Y;Z)-I(Y;Z|X), \\
 I(Y;X) +I(Z;X) - I(\{Y,Z\};X) = I(Y;X) - I(Y;X|Z)=I(Z;X)-I(Z;X|Y).
\end{eqnarray}

These identities taken together show that the definition of $R$ is symmetric under variable permutations and in particular proves our proposition
\begin{eqnarray}
R= I(X;Z) +I(Y;Z) - I(\{X,Y\};Z) =  I(X;Y) - I(X;Y|Z).
 \end{eqnarray}
 
As a corollary we see that $R$ is fully symmetric, i.e. 
  \begin{eqnarray} 
 R &&  = I(X;Z) +I(Y;Z) - I(\{X,Y\};Z) = I(X;Y) +I(Z;Y) - I(\{X,Z\};Y)  \nonumber \\
 && \qquad \qquad \qquad \qquad \qquad  \qquad \qquad \qquad = I(Y;X) +I(X;Z) - I(\{Y,Z\};X) \nonumber \\
 && = I(X;Y) - I(X;Y|Z) = I(X;Z)-I(X;Z|Y)=I(Y;Z)-I(Y;Z|X).
 \end{eqnarray}
 Thus $R$ can  be computed efficiently through the most convenient set of estimated distributions. 

In the case of more than three variables analogous expression for $R$ in terms of conditional mutual informations can be written easily, through (\ref{Iexpansion}), but become less elegant as the distributions are now conditional on several variables.

 \end{document}